\documentclass[11pt]{article}
\usepackage{moriond,epsfig}

\def\Journal#1#2#3#4{{#1} {\bf #2}, #3 (#4)}

\def\NPA{{\em Nucl. Phys.} A}

\def\PLB{{\em Phys. Lett.}  B}
\def\PRL{\em Phys. Rev. Lett.}
\def\PRC{{\em Phys. Rev.} C}
\def\PRD{{\em Phys. Rev.} D}

\def\be{\begin{equation}}
\def\ee{\end{equation}}
\def\bea{\begin{eqnarray}}
\def\eea{\end{eqnarray}}

\def\xe{$^{136}$Xe}
\newcommand{\PerTonDay}{(ton$\cdot$day)$^{-1}$}

\begin{document}
\vspace*{4cm}
\title{FIRST RESULT FROM KamLAND-Zen\\Double Beta Decay with \xe}

\author{ A. GANDO\\for the KamLAND-Zen Collaboration}

\address{Research Center for Neutrino Science, Tohoku University, Sendai 980-8578, Japan}

\maketitle\abstracts{
We present the first result of the KamLAND-Zen experiment, measurement of {\xe} double-beta decay. In an exposure of 77.6~days with 129~kg of {\xe}, two-neutrino double-beta decay half-life is measured precisely to be $T_{1/2}^{2\nu} = 2.38 \pm 0.02({\rm stat}) \pm 0.14({\rm syst}) \times 10^{21}$~yr.  This value is consistent with the measurement by \mbox{EXO-200} and significantly below the lower limit obtained by previous experiment.  For the neutrinoless double-beta decay half-life, improved lower limit is set to $T_{1/2}^{0\nu} > 5.7 \times 10^{24}$~yr at 90\% C.L. and corresponding upper limit of the effective neutrino mass is ranged to 0.3-0.6 eV depending on the adopted nuclear matrix elements.
}

\section{Introduction}
Double-beta decay is a very slow nuclear transition and there are two main modes of the decay. Two neutrino double-beta~(2$\nu\beta\beta$) decay which emits two electrons and two anti-neutrinos is described by known physics and have been observed with various nuclei. On the other hand, neutrinoless double-beta~(0$\nu\beta\beta$) decay which violates the lepton number conservation, is predicted theoretically, but not yet observed. It requires two characteristic neutrino properties; neutrinos have mass and are Majorana type lepton \cite{Schechter1982}. Although absolute mass of neutrinos are not known yet, neutrino oscillation experiments established that neutrinos have non zero mass. If 0$\nu\beta\beta$ decay is observed, we can conclude the neutrino is Majorana particle. In addition, if 0$\nu\beta\beta$ transition mechanism is simply exchange of a light Majorana neutrino, effective neutrino mass, $\left<m_{\beta\beta}\right> \equiv \left| \Sigma_{i} U_{ei}^{2}m_{\nu_{i}} \right|$, which is proportional to square root of 0$\nu\beta\beta$ decay rate, provide neutrino mass hierarchy and information of absolute neutrino mass scale. Allowed region of $\left<m_{\beta\beta}\right>$ narrowed by oscillation experiments is divided into three; degenerated hierarchy (m$_1 \sim$ m$_2 \sim$ m$_3$) where $\left<m_{\beta\beta}\right>$ is more than 50 meV, inverted hierarchy ($m_2 > m_1 \gg m_3$) ranging from 20 to 50 meV, and normal hierarchy ($m_3 \gg m_2 > m_1$) less than 20 meV~\cite{Nakamura2010}. In addition, there is a claim of observation of 0$\nu\beta\beta$ decay at a few hundred meV with $^{76}$Ge experiment (KKDC claim)~\cite{KKDC2006}. From an experimental point of view, features of double-beta decay, very long half-life ($>$10$^{18}$ yr) and a few-MeV $Q$ value, require a large amount of isotopes and low background environment. Recent experiments plan to use more than 100 kg of double-beta decay isotopes with various techniques to cover the degenerated hierarchy. Extension of nucleus mass to order of ton will enable to investigate and mostly cover the inverted hierarchy.

With large and clean environment of KamLAND, KamLAND-Zen~(KamLAND ZEro Neutrino double-beta decay) experiment studies 2$\nu\beta\beta$ and 0$\nu\beta\beta$ decays with {\xe}. The first phase was started in September, 2011, with $\sim$300 kg of {\xe} (the largest  isotope mass in current $\beta\beta$ decay experiments), to explore the degenerated hierarchy and test the KKDC claim.

\section{Detector}
\begin{figure}
\begin{center}
\includegraphics[bb=60 25 600 450,clip,width=0.6\hsize]{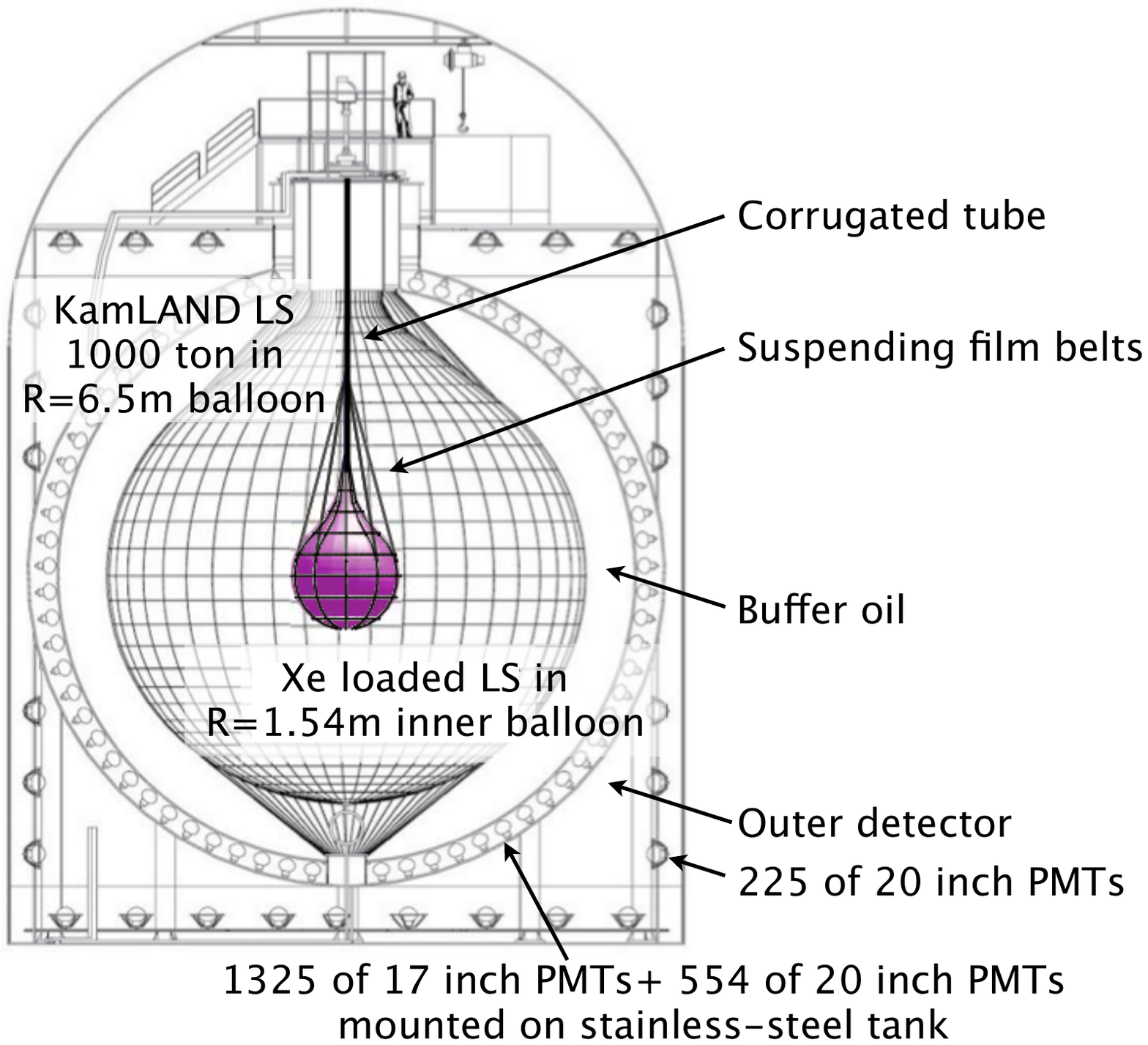}
\caption{Schematic view of KamLAND-Zen
\label{fig:detector}}
\end{center}
\end{figure}

KamLAND is a low energy anti-neutrino detector with 1 kton of highly purified liquid scintillator (LS) and KamLAND-Zen is a modification of the detector (Fig.~\ref{fig:detector}). Double beta decay isotope of KamLAND-Zen is $\sim$300 kg of {\xe} whose $Q$ value is 2.476 MeV. Advantages of using Xe in our experiment are soluble to LS $\sim$3\% by weight and easily extracted. Furthermore, isotopic enrichment and purification method are already established. The primary source/detector for double-beta decay is \mbox{13~tons} of Xe-loaded liquid scintillator~(Xe-LS) contained in a 1.54-m-radius spherical inner balloon (IB), suspended at the center of the detector. IB is made of 25-$\mu$m-thick transparent clean nylon film. Twelve belts made by the same material as IB and connected Vectran strings support the IB. Xe-LS is surrounded by 1 kton of LS contained in a 6.5-m-radius spherical balloon. The Xe-LS consists of 82\% decane and 18\% pseudocumene (1,2,4-trimethylbenzene) by volume, 2.7 g/liter of the fluor PPO (2,5-diphenyloxazole), and $(2.52 \pm 0.07)$~\% by weight of enriched xenon gas, as measured by gas chromatography. The isotopic abundances in the enriched xenon were measured by a residual gas analyzer to be $(90.93 \pm 0.05)\%$ \mbox{\xe}, $(8.89 \pm 0.01)\%$ \mbox{$^{134}$Xe}, and the other xenon isotopes are negligible. Density difference between Xe-LS and LS is controlled within 0.10\% to reduce the load for IB, and LS acts as an active shield for external $\gamma$'s and as a detector for internal radiation from the Xe-LS or IB. Scintillation light is monitored by 1,325 of 17-inch and 554 of 20-inch photomultiplier tubes (PMTs) mounted on the 18-m-diameter spherical stainless-steel tank. Stainless-steel tank is surrounded by outer detector fulfilled with 3.2~kton of pure water as a Cherenkov detector. 225 of 20-inch PMTs are mounted on the wall. Detail of KamLAND detector is described in Ref.~\cite{Abe2010}.

Modification of KamLAND for the double beta decay experiment was done in August, 2011. The most important work, inner balloon installation was carried out by nearly 20 researchers and students. Since balloon and corrugated tube has $\sim$14-m-long in order to put that at the detector center, people hold the folded balloon and slowly forward with repetition of sending small amount of liquid to forefront of inner balloon to sink it.  Installation successfully finished within 2 hours. IB in the LS was monitored by ``monitor system" consist of web cameras and LED lights. Inner balloon was inflated with ``dummy-LS" which is Xe-free LS, to check its shape and find possibility of leakage by small amount of Rn as a tracer contained in a LS pipe line. After that, we started Xe-LS filling replacing with dummy-LS. Thanks to KamLAND, whose DAQ and analysis tools were already established, we took data to monitor the inner balloon condition during the inflation and Xe-LS filling. Data taking for double beta decay started on September 24, 2011. After Rn decay and removal of ``monitor system'', the data presented here were collected between October 12, 2011, and January 2, 2012.

\section{Calibration and systematic uncertainty}
\begin{figure}
\begin{minipage}{0.5\hsize}
\begin{flushleft}
\includegraphics[width=\hsize]{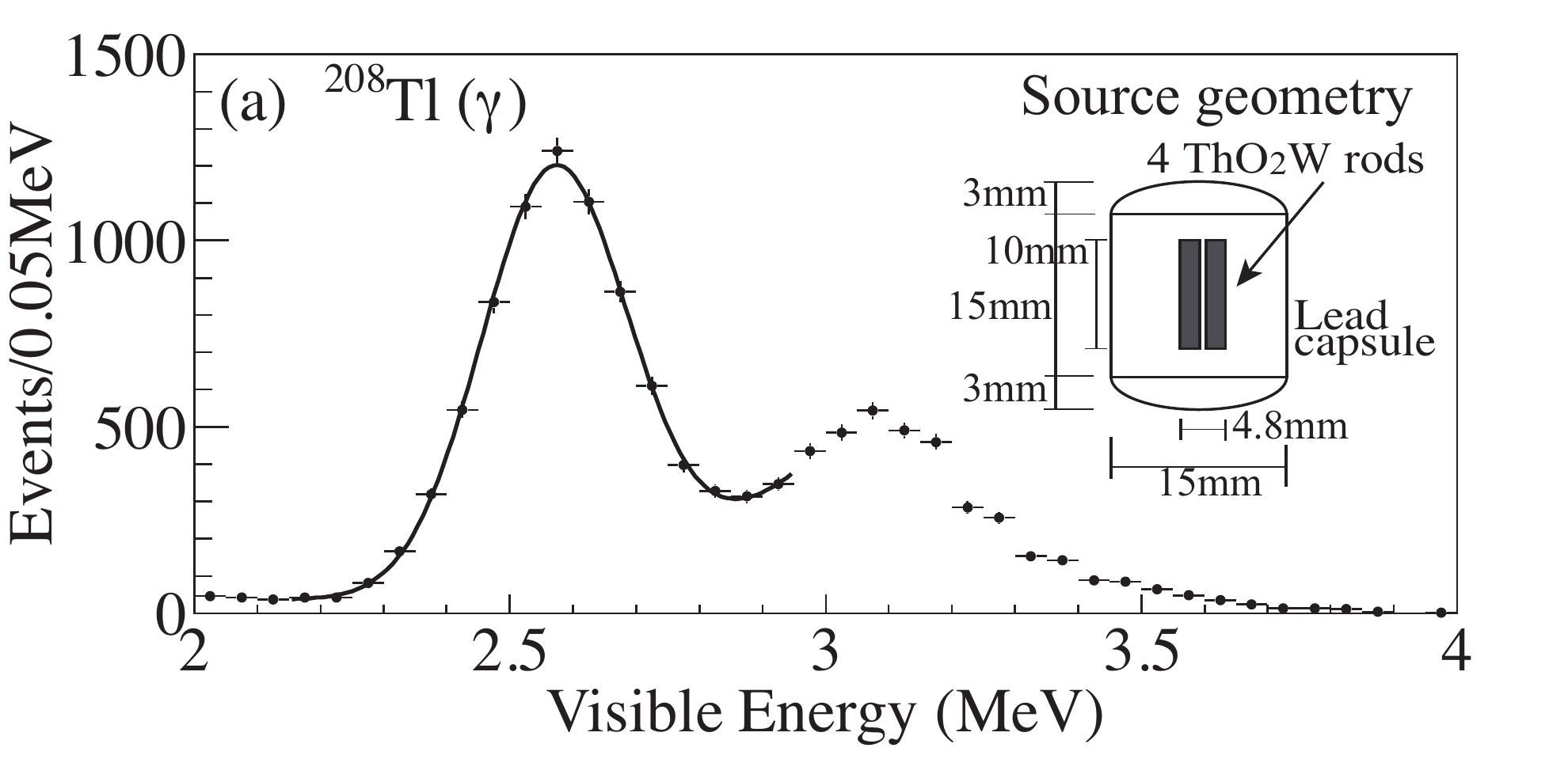}
\includegraphics[width=\hsize]{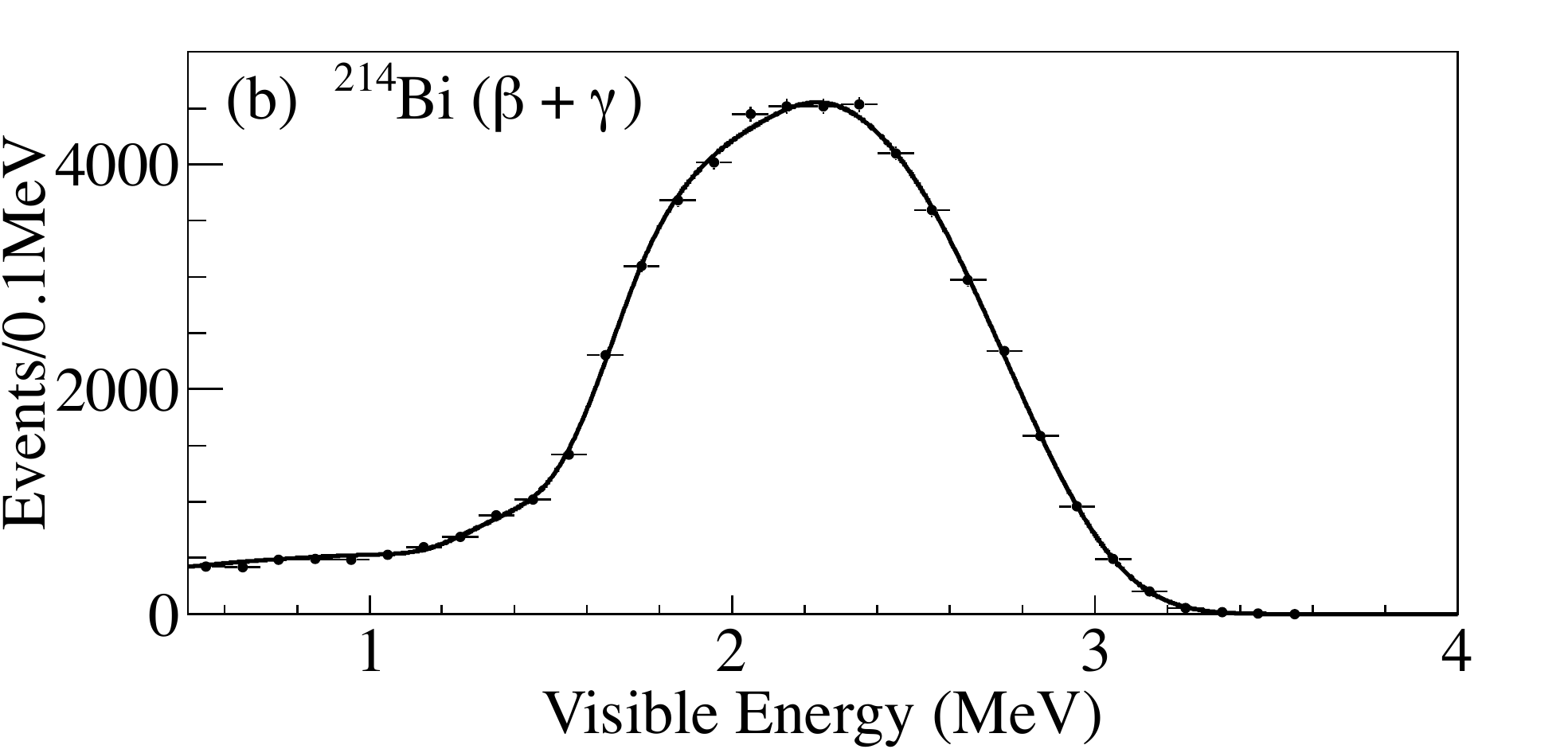}
\end{flushleft}
\end{minipage}
\begin{minipage}{0.5\hsize}
\begin{flushright}
\includegraphics[width=\hsize]{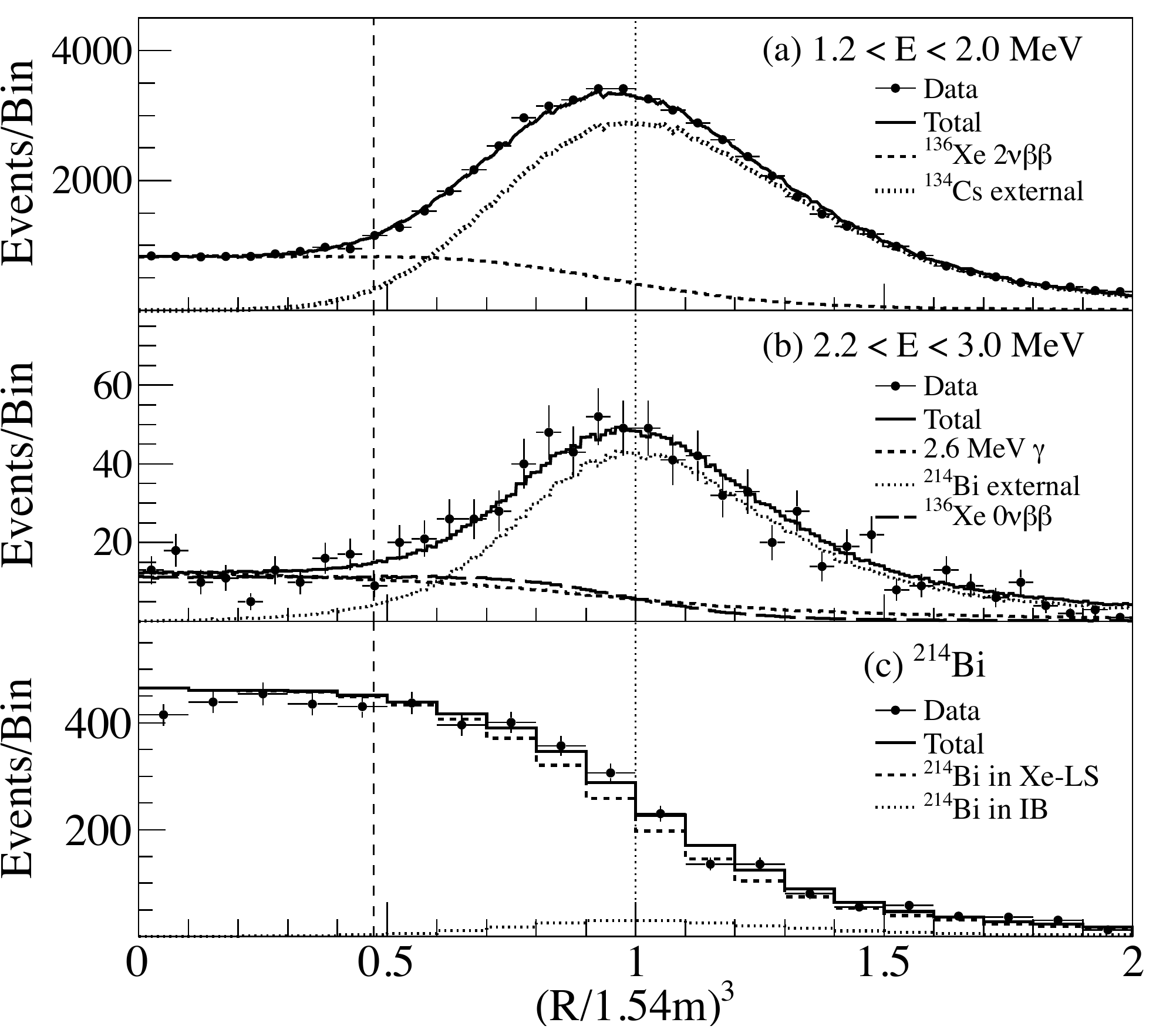}
\end{flushright}
\end{minipage}
\begin{flushleft}
\caption{[Left] Visible energy distribution of (a) $\gamma$'s from the ThO$_2$-W ($^{208}$Tl) source and (b) $^{214}$Bi ($\beta + \gamma$) decays in the Xe-LS. The line indicate the best-fit curves. The fit to (a) has a $\chi^2$/d.o.f. = 5.0/8 and (b) has a $\chi^2$/d.o.f. = 27.0/29. In (a), geometry of calibration source is also shown.
[Right] R$^3$ distribution of candidate events for (a) $1.2 < E < 2.0$ MeV ($^{136}$Xe 2$\nu\beta\beta$ window), (b) $2.2 < E < 3.0$ MeV (0$\nu\beta\beta$ window) and (c) $^{214}$Bi. Lines indicate the best-fit components. (a) {\xe} 2$\nu\beta\beta$ (dashed) and $^{134}$Cs (dotted). (b) 2.6 MeV $\gamma$ (dashed) and $^{214}$Bi (dotted); {\xe} 0$\nu\beta\beta$ (long-dashed) is also shown instead of 2.6 MeV $\gamma$. (c) $^{214}$Bi from Xe-LS (dashed) and $^{214}$Bi from IB (dotted). The vertical lines show the fiducial 1.2-m-radius (dashed) and IB radius (dotted).~
\label{fig:attenuation}}
\end{flushleft}
\end{figure}

KamLAND-Zen observes the summed energy of two electrons in the double-beta decay due to inseparable scintillation light of those.
Event energy (visible energy) is estimated by the number of PMT hit and observed photoelectron (p.e.) after correcting various parameters such as gain, dark rate and transparency depending on the event vertices.
The energy response is calibrated with three kind of sources that (1) $^{208}$Tl 2.614 MeV $\gamma$'s from artificial calibration source of ThO$_2$-W, (2) 2.225 MeV $\gamma$'s from spallation neutrons capture on protons, and (3) $^{214}$Bi ($\beta + \gamma$) from $^{222}$Rn ($\tau$ = 5.5 days) introduced during filling of liquid scintillator. From the energy distribution of (1) and Monte Carlo study, energy resolution is estimated to be $\sigma$ = (6.6 $\pm$ 0.3)\%/$\sqrt{E({\rm MeV})}$. The systematic variation of the energy reconstruction is monitored by (2) and it is less than 1.0\%. The detector energy response is also stable within 1.0\%. Energy nonlinearity effects caused by scintillator quenching and Cherenkov light production are constrained by (1) 2.614 MeV peak of $^{208}$Tl $\gamma$ (Fig.\ref{fig:attenuation} [Left], (a)) and spectrum shape of (3) $^{214}$Bi, fitted well with simulation data (Fig.\ref{fig:attenuation} [Left], (b)). The vertex resolution is estimated to be $\sigma\sim$15~cm/$\sqrt{\rm {E(MeV}}$) from radial vertex distribution of $1.2 < E < 2.0$ MeV and $2.2 < E < 3.0$ MeV window (Fig.~\ref{fig:attenuation} [Right], (a) and (b)). Fig.~\ref{fig:attenuation} [Right] shows the event density variation in the Xe-LS as a function of the distance from IB center. The vertex reconstruction performance is estimated with that of $^{214}$Bi (Fig.~\ref{fig:attenuation} [Right], (c)).

Fiducial volume (1.2-m-radius) is selected to mitigate the background from IB material, seen in Figures~\ref{fig:attenuation} [Right], (a) and (b). Xe-LS volume in the fiducial volume is 7.24 m$^3$, corresponding amount of {\xe} is 129 kg, and volume ratio of FV to total volume (16.51 $\pm$ 0.17 m$^3$) is 0.438 $\pm$ 0.005. Total volume was measured by a flow meter during the deployment. This volume fraction is compared with event-counting ratio of $^{214}$Bi. In this calculation, IB surface background contribution is subtracted and its uncertainty is included to the event-counting ratio. The total difference of those ratio gives an estimate of the FV uncertainty (5.2\%). Other systematic uncertainties come from enrichment of {\xe} (0.05\%), Xe concentration (2.8\%), detector energy scale (0.3\%), Xe-LS edge effect (0.06\%), and detection efficiency (0.2\%), and total systematic uncertainty is evaluated to 5.9\% from the quadrature sum of individual contributions.

\section{Background}
Main background of measurement comes from (1) radioactive impurities from inner balloon (external), (2) those from Xe-LS and (3) spallation products generated by cosmic-ray muons. Originally, (1) was thought to be dominated by $^{238}$U and $^{232}$Th chain and $^{40}$K, however, from radial vertex distribution, it reveals that there are unexpected backgrounds in 2$\nu\beta\beta$ energy region. Expected background with (2) is only $^{238}$U and $^{232}$Th chain, $^{85}$Kr and $^{210}$Bi but unknown peak appear in the 0$\nu\beta\beta$ window ($2.2 < E < 3.0$ MeV). For (3), dominated spallation product from $^{12}$C in the $^{136}$Xe double-beta decay energy range is $^{10}$C ($\beta^{+}$ decay, \mbox{$\tau = 27.8$~s}, \mbox{$Q = 3.65$~MeV}) and $^{11}$C ($\beta^{+}$, \mbox{$\tau=29.4$~min}, \mbox{$Q=1.98$~MeV}). The spallation product from xenon is also estimated with lifetime $<$ 100 sec and is revealed to be small.

Unexpected peak in the 0$\nu\beta\beta$ window is fitted with 0$\nu\beta\beta$ spectrum firstly (see Fig.~\ref{fig:energy} (c)) but found to have $\sim$3\% energy difference, and the hypothesis that the peak is explained by $0\nu\beta\beta$ event only is excluded at more than 5$\sigma$ C.L. by $\chi^2$ test. Events in the 0$\nu\beta\beta$ window have stable rate and distribute uniformly in Xe-LS. The well known 2.614 MeV gamma from $^{208}$Tl ($\beta^-$ decay, $Q$ = 5.0 MeV) as a candidate 0$\nu\beta\beta$ background, is actually distributed to 3-5 MeV due to the coincident $\beta$/$\gamma$ detection in the surrounding LS. There are two possibilities of the background; long-lived radioactive impurities or cosmogenic spallation nuclei. Unfortunately, identification of $\beta$ or $\gamma$ from the small difference of those radial distributions is difficult, and ex-situ measurements don't have enough sensitivity to determine isotopes due to its small amount. So we search all isotopes and decay path in ENSDF~\cite{ENSDF2006} database as the following procedure. For all tabulated decay chains in the database, the visible energy spectrum is calculated taking into account of alpha quenching, energy resolution, energy non-linearity and the time structure of the chain and pile-up in DAQ. Considering only nuclei with peak structure is in 2.4-2.8 visible-MeV, and has long-lived parent whose lifetime is longer than 30 days, only 4 nuclei remained; $^{\rm 110}$Ag$^m$ ($\beta^{-}$ decay, \mbox{$\tau=360$~day}, \mbox{$Q = 3.01$~MeV}), $^{88}$Y (EC decay, \mbox{$\tau=154$~day}, \mbox{$Q = 3.62$~MeV}), $^{208}$Bi (EC decay, \mbox{$\tau=5.31 \times 10^{5}$~yr}, \mbox{$Q = 2.88$~MeV}), and $^{60}$Co ($\beta^{-}$ decay, \mbox{$\tau = 7.61$~yr}, \mbox{$Q = 2.82$~MeV}) as potential background sources. Although $^{88}$Y and $^{60}$Co is a little bit constrained by its half-life and shape, remained 4 background candidates are all included in the fitting as a free parameters since there is no clear identification of the background nor determination by an ex-situ measurement. Presence of $^{110}$Ag$^m$ may be explained by the spallation of {\xe} at aboveground caused by high cosmic ray flux since Xe gas was enriched in Russia and sent to Japan by airplane, or fallout by the Fukushima I reactor accident in March 11, 2011 since it is observed with Ge detector in the soil sample in Sendai where is located 100 km away from the reactor. The possibility of $^{208}$Bi is difficult to explain. From Xe gas bottle and filter of Xe system used for making Xe-LS, $^{209}$Bi, not $^{208}$Bi, was detected with finite value by ICP-MS, but for Xe-LS, it was less than detection sensitivity. Calculated amount of $^{208}$Bi from measured $^{209}$Bi in Xe-LS cannot explain the peak in 0$\nu\beta\beta$ window. If bismuth was contained in Xe-LS, the 2.3 MeV peak of $^{207}$Bi should also appear but no clear peak found and ratio of $^{207}$Bi/$^{208}$Bi is much smaller than expected. Short-lived nuclei with a lifetime between 100 sec and 30 days possibly supported by muon spallation is stringently constrained from the study of energy spectrum with close (A, Z) nuclei scaled with the production cross sections~\cite{Napolitani2007}. The estimated value from data with lifetime $<$ 100 sec associated with muons depositing more than $\sim$3 GeV is 0.02~\PerTonDay\, at 90\% C.L., where ton is a unit of the Xe-LS mass.

Unexpected background also existed in the 2$\nu\beta\beta$ decay region. $^{134}$Cs ($\beta$ + $\gamma$) is dominating around the IB boundary. This rare nuclei not existing in nature is brought by the fallout of Fukushima I reactor since fabrication of IB was done in Sendai, and ratio of observed activity of $^{134}$Cs/$^{137}$Cs is consistent with soil sample measurements. Amount of $^{137}$Cs can't be explained by the spallation of {\xe} as Cs is not observed in Xe-LS. Visible energy distribution as a function of R$^3$ distribution also show the two peak of $^{134}$Cs on the IB.

\section{Result}
We measured the half-lives of {\xe} $2\nu\beta\beta$ and $0\nu\beta\beta$ decays from likelihood fit to the binned energy distribution between 0.5 and 4.8 MeV. The 2$\nu\beta\beta$, 0$\nu\beta\beta$ decays, and backgrounds are fitted simultaneously. Main backgrounds contained in Xe-LS such as $^{85}$Kr, $^{40}$K, $^{210}$Bi and the $^{238}$U-$^{222}$Rn and $^{232}$Th-$^{224}$Ra decay chains are unconstrained, while $^{222}$Rn-$^{210}$Pb and $^{228}$Th-$^{208}$Pb chains, $^{10}$C, $^{11}$C and contribution from IB such as $^{238}$U chain, $^{232}$Th chain, $^{134}$Cs and $^{137}$Cs are allowed to vary but constrained by the estimated rates respectively. For the possible backgrounds in Xe-LS, we also include reactor fallout which was found in the ex-situ measurement and have half-life longer than 30 days, such as $^{137}$Cs, $^{134}$Cs, $^{\rm 110}$Ag$^m$, $^{\rm129}$Te$^m$, $^{95}$Nb, $^{90}$Y (from $^{90}$Sr), and $^{89}$Sr in the fit without any constraint. For the 0$\nu\beta\beta$ window, 4 nuclei ($^{\rm 110}$Ag$^m$, $^{208}$Bi, $^{88}$Y, $^{60}$Co) searched from ENSDF are included in the fit as unconstrained free parameters. Half-lives of each nuclei is considered with $\chi^2_{time}$. Calibration source $^{208}$Tl and radon-induced $^{214}$Bi are used for constraint of the energy scale uncertainty.

\begin{figure}
\begin{minipage}{0.5\hsize}
\begin{flushleft}
\includegraphics[height=2.8in]{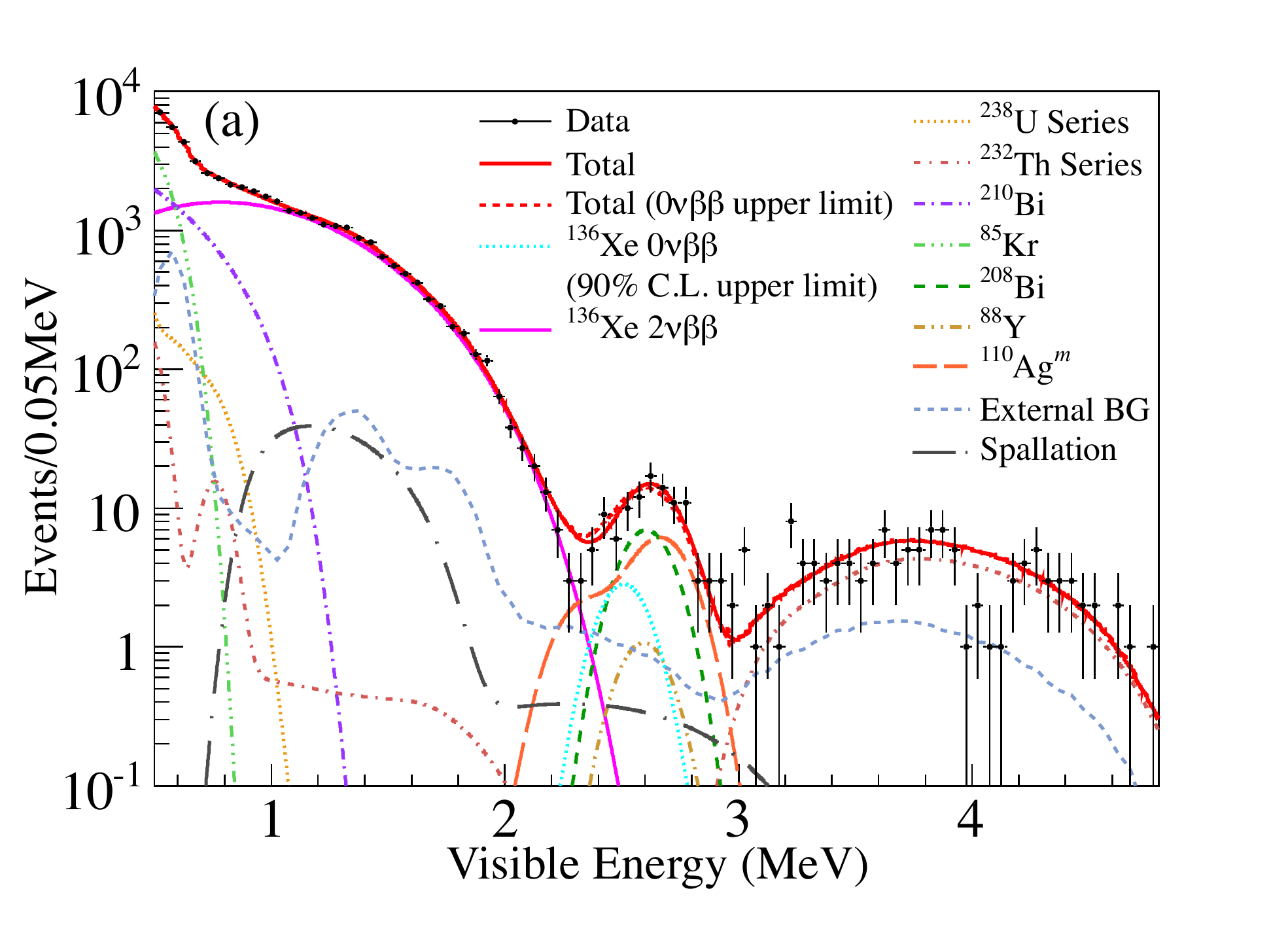}
\end{flushleft}
\end{minipage}
\begin{minipage}{0.5\hsize}
\begin{flushright}
\includegraphics[height=3.5in]{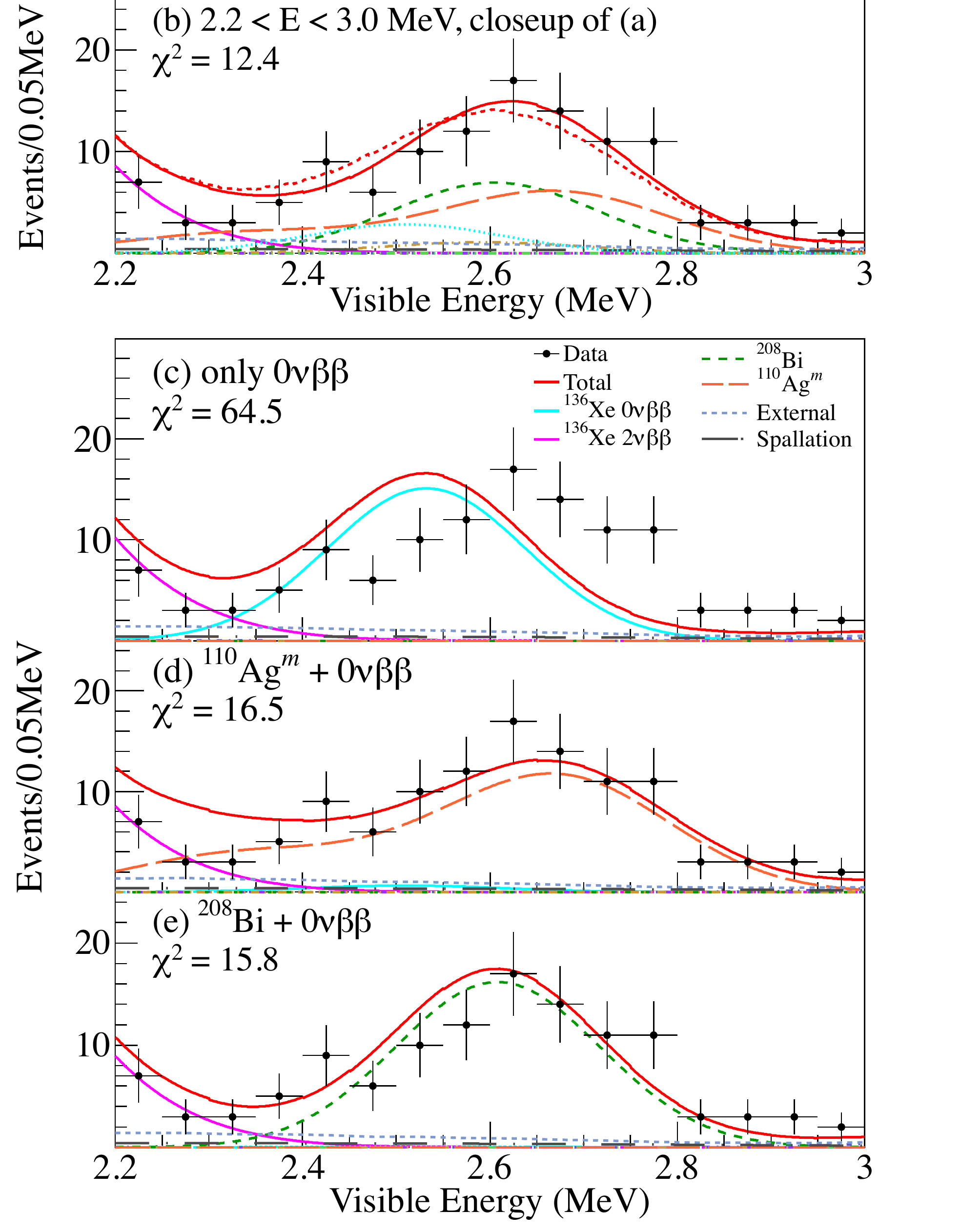}
\end{flushright}
\end{minipage}
\caption{(a) Energy spectrum of {\xe} $2\nu\beta\beta$ decay and best fit backgrounds in $0.5 < E < 4.8$ MeV. 90\% C.L. upper limit of $0\nu\beta\beta$ decay is also shown. (b) Close up of (a) for $2.2 < E < 3.0$ MeV. (c)-(e) Close up for $2.2 < E < 3.0$ MeV with assumption of (c) only 0$\nu\beta\beta$, (d) $^{\rm 110}$Ag$^m$ + 0$\nu\beta\beta$, (e) $^{208}$Bi+ 0$\nu\beta\beta$ existed, for comparison. $\chi^2$ provided in (b)-(e) is taken from data in shown energy range.
\label{fig:energy}}
\end{figure}

The result of fit is presented in the Figure~\ref{fig:energy}(a) and it has a $\chi^2$/d.o.f. = 99.7/87 obtained by comparing the binned data and the best fit expectation. Totally 80\% of {\xe} 2$\nu\beta\beta$ spectrum is in the fitting range and the best-fit number of decay is $(3.55 \pm 0.03) \times 10^{4}$, which correspond to an event rate of $80.9 \pm 0.7$~\PerTonDay. Combined background decay rate of $^{\rm 110}$Ag$^m$, $^{208}$Bi, $^{88}$Y and $^{60}$Co is $0.22 \pm 0.04$~\PerTonDay. In Figures~\ref{fig:energy}(d) and \ref{fig:energy}(e), fitting result with an assumption that only one kind of nuclei among 4 kinds found in ENSDF is relevant to the background is shown for $^{\rm 110}$Ag$^m$ and $^{208}$Bi. $0\nu\beta\beta$ limit is calculated by floating these backgrounds [Fig.~\ref{fig:energy}(b)]. The upper limit (90\% C.L.) of $0\nu\beta\beta$ decay is $<$ 15~events, and corresponding event rate is $<$ 0.034~\PerTonDay. Decay rate of 2$\nu\beta\beta$ and event rate in 0$\nu\beta\beta$ window ($2.2 < E < 3.0$ MeV) are stable during the data-set.
 
From the fitting result, the measured half-life of {\xe} $2\nu\beta\beta$ decay is $T_{1/2}^{2\nu} = 2.38 \pm 0.02({\rm stat}) \pm 0.14({\rm syst}) \times 10^{21}$~yr, which is consistent with EXO-200 result, $T_{1/2}^{2\nu} = 2.11 \pm 0.04({\rm stat}) \pm 0.21({\rm syst}) \times 10^{21}$~yr~\cite{Ackerman2011} and fall below the lower limit, $T_{1/2}^{2\nu} > 1.0 \times 10^{22}$~yr, obtained by DAMA~\cite{Bernabei2002}. Estimated lower limit of {\xe} $0\nu\beta\beta$ decay half-life is $T_{1/2}^{0\nu} > 5.7 \times 10^{24}$~yr at 90\% C.L.. This corresponds to five-fold improvement over previous limits~\cite{Bernabei2002}. From this limit, corresponding upper limit of effective neutrino mass via QRPA (CCM SRC)~\cite{Simkovic2009} and shell model~\cite{Menendez2009} is 0.3-0.6 eV.

\section{Summary}
KamLAND-Zen measured {\xe} 2$\nu\beta\beta$ decay half-life the most precisely. This result is consistent with the measurement by \mbox{EXO-200} and significantly below the lower limit obtained by a previous experiment. Obtained lower limit of 0$\nu\beta\beta$ decay half-life is almost fivefold improvement over previous experiments. To improve the sensitivity of 0$\nu\beta\beta$ decay search, removal of contaminants in the Xe-LS such as distillation is planned. In the future, the largest systematic uncertainty from fiducial volume determination will be reduced by source calibration inside of the inner balloon.

\section*{Acknowledgments}
Talk and this manuscript is based on the paper~\cite{KamLAND-Zen2012}. The KamLAND-Zen experiment is supported by the Japanese Ministry of Education, Culture, Sports, Science and Technology (MEXT) and the US Department of Energy (DOE). The Kamioka Mining and Smelting Company has provided service for activities in the mine.

\end{document}